# Contrôle paracrine du développement du tissu adipeux par l'autotaxine et l'acide lysophosphatidique

Jean Sébastien Saulnier-Blache.

IFR31 Institut Louis Bugnard, BP 84225, INSERM U586, Unité de Recherches sur les Obésités, 31432 TOULOUSE Cedex 4 – France. saulnier@toulouse.inserm.fr

Abstract : Secretion and role of autotaxin and lysophosphatidic acid in adipose tissue

In obesity, adipocyte hypertrophy is often associated with recrutement of new fat cells (adipogenesis) under the control of circulating and local regulatory factors. Among the different lipids released in the extracellular compartment of adipocytes, our group found the presence of lysophosphatidic acid (LPA). LPA is a bioactive phospholipid able to regulate several cell responses via the activation of specific G-protein coupled membrane receptors. Our group found that LPA increases preadipocyte proliferation and inhibits adipogenesis via the activation of LPA1 receptor subtype. Extracellular LPA-synthesis is catalyzed by a lysophospholipase D secreted by adipocytes : autotaxin (ATX). Adipocyte ATX expression strongly increases with adipogenesis as well as in individuals exhibiting type 2 diabetes associated with massive obesity. A possible contribution of ATX and LPA as paracrine regulators of adipogenesis and obesity associated diabetes is proposed.

Résumé : Dans l'obésité, l'hypertrophie adipocytaire s'accompagne souvent du recrutement de nouveaux adipocytes, ou adipogénèse. Cet événement cellulaire, est sous la dépendance d'hormones ainsi que de facteurs sécrétés au sein même du tissu adipeux. Parmi les lipides produits dans le compartiment extracellulaire des adipocytes, notre groupe a montré la présence d'acide lysophosphatidique (LPA). Le LPA est un phospholipide bioactif capable de réguler plusieurs réponses cellulaires via l'activation de récepteurs membranaires spécifiques couplés aux protéines G. Nous avons montré que le LPA, via l'activation du récepteur LPA1, augmentait la prolifération des préadipocytes et inhibait l'adipogénèse. La synthèse extracellulaire de LPA est catalysée par une lysophospholipase D sécrétée par l'adipocyte : l'autotaxine (ATX). L'expression de l'ATX adipocytaire augmente au cours de l'adipogénèse ainsi que dans le tissu adipeux d'individus présentant un diabète de type 2 associé à une obésité massive. Un rôle du LPA et le l'ATX comme régulateurs paracrine de l'adipogénèse et/ou du diabète associé à l'obésité est donc envisagé.

En raison de la gravité des pathologies qui lui sont associées (hypertension artérielle, diabète, athérosclérose, cancer), l'obésité constitue un problème majeur de santé publique dans les sociétés industrialisées. L'amélioration de la prise en charge et de la prévention de l'obésité requiert une bonne connaissance des mécanismes de régulation du métabolisme et du développement du tissu adipeux, organe clé de cette pathologie.

L'obésité se caractérise par une augmentation excessive de la masse du tissu adipeux associée à une augmentation des réserves de triglycérides accumulés dans les adipocytes (hypertrophie). L'hypertrophie adipocytaire s'accompagne fréquemment du recrutement de nouveaux adipocytes (hyperplasie) issus d'un processus de différentiation (adipogénèse) de précurseurs adipocytaires dormants (préadipocytes) présents tout au long de la vie dans l'environnement immédiat des adipocytes. L'identification des facteurs impliqués dans la régulation de l'adipogénèse est crucial pour envisager un contrôle de l'hyperplasie adipocytaire. Certains de ces facteurs (insuline, catécholamines, hormones thyroïdiennes ou stéroïdes, …) sont exogènes au tissu adipeux et apportés par voie sanguine. D'autres facteurs (cytokines, prostanoïdes, phospholipides, …) sont produits au sein même du tissu adipeux où ils agissent de façon autocrines et/ou paracrine.

Cette revue synthétise les connaissances actuelles sur un des facteurs produit localement dans le tissu adipeux, l'acide lysophosphatidique (LPA) et l'enzyme impliquée dans sa synthèse : l'autotaxine (ATX).

**1/ Production de LPA dans le milieu extracellulaire des adipocytes.**

Plusieurs études ont montré que des milieux de culture conditionnés par des d'adipocytes ont la capacité d'augmenter la prolifération de préadipocytes en culture auxquels ils sont exposés (Shillabeer et al 1989 ; Considine et al 1996). Nous avons montré qu'une partie de cet effet prolifératif pouvait être bloqué par un pré-traitement de ces milieux conditionnés par une enzyme hydrolysant spécifiquement les lysophospholipides : la phospholipase B (Valet et al 1998). Une analyse qualitative (basée sur le pré-marquage des phospholipides cellulaires au [$^{32}$P]) du contenu en phospholipides des milieux conditionnés, a révélé la présence de plusieurs lysophospholipides, en particulier l'acide lysophosphatidique (LPA) (Valet et al 1998 ; Gesta et al 2002). La mise au point d'un dosage radioenzymatique spécifique du LPA a permis de confirmer sa présence et de le quantifier (Saulnier-Blache et al. 2000).

Le LPA est un phospholipide bioactif connu pour réguler certains évènements cellulaires (prolifération, migration, apoptose...) via l'activation de récepteurs membranaires spécifiques couplés aux protéines G hétérotrimériques. A ce jour, quatre récepteurs au LPA ont été identifiés et clonés : LPA-1,-2,-3,-4 (Anliker et al 2004). Parallèlement, bien que controversées, certaines études ont proposé que le LPA pouvait également être un ligand d'un facteur de transcription connu pour son rôle crucial dans le processus d'adipogénèse : PPARγ (McIntyre et al 2002).

Le LPA est également présent dans le plasma (Saulnier-Blache et al. 2000). Cependant il n'existe aucune corrélation entre les taux plasmatiques de LPA et l'état de masse grasse chez l'animal ou chez l'Homme. Ceci montre que le LPA produit par le tissu adipeux ne semble pas contribuer de façon significative aux taux circulants de LPA. Ainsi, la production LPA par le tissu adipeux semble être principalement de nature locale, suggérant des effets autocrines et/ou paracrines plutôt qu'endocrine.

**2/ Les récepteurs au LPA dans le tissu adipeux**

Le LPA étant connu pour agir via l'activation de récepteurs spécifiques, leur présence dans le tissu adipeux a été étudiée. Sur la base de la quantification des ARNm, le récepteur

LPA1 est apparu comme le sous-type le plus abondant dans le tissu adipeux (humain et murin). Le sous-type LPA2 est également présent, mais de façon très minoritaire. Les récepteurs LPA3 et LPA4 sont absents (Pagès et al 2001 ; Simon et al 2005). Bien que le facteur de transcription PPARγ2 soit abondamment exprimé dans l'adipocyte, l'activation de PPARγ2 par le LPA n'a pas pu être mis en évidence dans l'adipocyte (Simon et al 2005). L'expression du récepteur LPA1 est 3 à 5 fois plus faible dans la fraction adipocytaire que dans la fraction stroma-vasculaire du tissu adipeux. Cette fraction contient des préadipocytes ainsi que des macrophages et des cellules endothéliales. L'expression relative du récepteur LPA1 dans les différents types cellulaires de la fraction stroma-vasculaire n'a pas été précisément déterminée. En culture, la différentiation des préadipocytes en adipocytes, s'accompagne d'une forte réduction de l'expression des ARNm du récepteur LPA1 (Pagès et al 2001; Simon et al 2005). Ainsi, via ce récepteur, le préadipocyte apparaît comme une des cibles possibles du LPA au sein du tissu adipeux.

**3/ Effets prolifératif et anti-adipogénique du LPA.**

L'expression du récepteur LPA1 dans les préadipocytes nous a conduit à tester l'action du LPA sur ces cellules. Des préadipocytes de la lignée murine 3T3F442A ou des préadipocytes en culture primaire issus de la fraction stroma-vasculaire de tissu adipeux, ont été traités par une des espèces moléculaires les plus actives de LPA : le 1-oléoyl-LPA (C18 :1). Ces études ont permis de confirmer que le LPA augmentait la croissance des préadipocytes (Valet et al 1998 ; Pagès et al 2001). De plus, le traitement au LPA inhibe la différenciation des préadipocytes en adipocytes (adipogénèse) (Simon et al 2005). Cet effet anti-adipogénique se caractérise par une inhibition de l'accumulation de triglycérides et une diminution d'expression de plusieurs gènes adipocytaires, parmi lesquels le facteur de transcription PPARγ2. Les préadipocytes isolés du tissu adipeux de souris invalidées pour le récepteur LPA1 possèdent une plus grande propension à se différencier en adipocytes que les préadipocytes de souris sauvages. De plus, l'effet anti-adipogénique du LPA disparaît après invalidation récepteur LPA1 (Simon et al 2005). Enfin, les souris invalidées pour le récepteur LPA1 présentent une augmentation significative du pourcentage de masse grasse comparativement aux souris sauvages (Simon et al 2005).

L'ensemble de ces observations suggèrent que le LPA (via son effet anti-adipogénique dépendant du récepteur LPA1) produit dans le milieu extracellulaire des adipocytes, pourrait exercer un effet inhibiteur local sur le recrutement de nouveau adipocytes au sein du tissu adipeux.

**4/ Implication d'une lysophospholipase D sécrétée dans la synthèse de LPA par l'adipocyte.**

Parallèlement à l'étude des effets biologiques du LPA il était important d'identifier les voies métaboliques impliquées dans sa synthèse. Au plan biochimique, plusieurs enzymes (phospholipases A, glycérolphosphate acyltransférase, monoacylglycérol kinase, lysophospholipases) sont susceptibles de catalyser la synthèse de LPA. Le LPA étant un phospholipide très polaire pouvant difficilement traverser la membrane plasmique, sa présence dans le compartiment extracellulaire semble devoir résulter de l'action d'une enzyme ectopique ou sécrétée, plutôt que d'une enzyme intracellulaire.

Nous avons observé que, après séparation des cellules, les concentrations de LPA dans les milieux conditionnés d'adipocytes n'étaient pas stables et augmentaient avec le temps (Gesta et al 2001). Cette observation suggérait l'existence d'une activité de synthèse du LPA (LPA-SA) sécrétée par les adipocytes. Une telle activité enzymatique avait déjà été décrite dans des plasma de rat (Tokumura et al *1986)*. La caractérisation biochimique de la LPA-SA adipocytaire a révélé qu'il s'agissait d'une activité de type lysophospholipase D (lysoPLD) soluble catalysant l'hydrolyse de lysophosphatidylcholine (LPC) (produite également en grande quantité par les adipocytes) en LPA et choline. Cette activité lysoPLD étant dépendante de la présence de métaux lourds (particulièrement le cobalt), elle peut être inhibé par la phénanthroline *(Gesta et al 2002).*

**4/ L'activité lysophospholipase D sécrétée par l'adipocyte est catalysée par l'autotaxine.**

A l'époque de ces observations, aucune enzyme capable de catalyser une activité lysoPLD sécrétée n'avait été identifiée. L'activité lysoPLD sécrétée dans les milieux de culture d'adipocytes a été purifiée, microséquencée et identifiée comme étant catalysée par l'autotaxine (ATX) *(Ferry et al 2003).*

L'ATX est une glycoprotéine de 100-110 kDa appartenant à la famille des ecto-nucleotide pyrophosphatase phosphodiestérase (enpp) (Gijsbers et al 2003). L'ATX a initialement été identifiée comme un puissant activateur de motilité cellulaire sécrété dans les milieux de culture d'une lignée de mélanome humain [*Stracke et al 1992*]. Cette enzyme a d'abord été décrite pour ses capacités d'hydrolyse des liaisons phosphodiesters de nucléotide comme l'ATP ou l'UDP-glucose [Clair & al. 1997]. Plus récemment, il a été démontré que l'ATX possédait également une activité de type lysophospholipase D (lyso-PLD) conduisant la synthèse de LPA ou de sphingosine-1-phosphate (un autre phospholipide bioactif) à partir de la lysophosphatidylcholine (LPC) ou de la la sphingosylphosphorylcholine (SPC) (Umezu-Goto & al. 2002 ; Clair et al 2003). Les activités PDE et lyso-PLD de l'ATX sont portées par le même site catalytique. Cependant, l'activité PDE semble être minoritaire par rapport à l'activité lyso-PLD, probablement en raison de l'accessibilité différentielle des substrats (LPC vs SPC) (Cimpean et al 2004).

Chez l'individu normal (Homme et souris), l'ATX est présente dans plusieurs tissus (cerveau, placenta, rein, pancréas, tissu adipeux) avec une expression plus importante dans le système nerveux central (Lee et al 1996 ; Kawagoe et al 1995), où elle pourrait être impliquée dans la neurogénèse et la myélinisation des oligodendrocytes [*Fox et al 2003*]. L'ATX est également présente au niveau plasmatique (Umezu-Goto et al 2002 ; Tokumura et al 2002), mais le type cellulaire et/ou le tissu à l'origine de cette ATX circulante reste inconnue.

L'expression de l'ATX est augmentée dans plusieurs types de cancers humains (carcinomes mammaires, hépatiques, thyroïdiens, neuroblastomes) [Yang et al 2002]. Les mécanismes impliqués dans ces augmentations restent inconnus. Par ailleurs, l'ATX a été décrite pour augmenter la tumorigenèse et l'angiogénèse *in vitro* et *in vivo* [Nam et al 2001]. Ces effets biologiques sont imputables à la synthèse de LPA et à l'activation de ses récepteurs. Les propriétés pro-métastasiantes et pro-angiogéniques l'ATX pourrait favoriser l'invasion des cellules cancéreuses et participer à la néovascularisation des tumeurs.

**5/ Régulations d'expression de l'ATX dans le tissu adipeux**

L'identification de l'ATX comme enzyme responsable de la synthèse de LPA dans le milieu extracellulaire des adipocytes, nous a conduit à rechercher les situations physiopathologiques associées à des modifications de son expression dans le tissu adipeux.

En culture, alors que l'expression (ARNm et protéine) de l'ATX est très faible (voire indétectable) dans les préadipocytes (lignée 3T3F442A ou culture primaire), elle augmente fortement au cours de la différenciation adipocytaire (Ferry et al 2003). Cette augmentation a également été observée au cours de la différenciation adipocytaire de cellules souches embryonaires de souris en culture (Ahn et al 2004). Cette augmentation ne semble pas directement attribuable au processus d'adipogénèse car, contrairement à d'autres gènes adipocytaires, l'expression de l'ATX est très fortement diminuée par à un traitement avec un ligand du facteur de transcription PPARγ. Inversement, le TNFα, cytokine connue pour son puissant effet anti-adipogénique, augmente l'expression de l'ATX (Boucher et al 2005). Les facteurs et les mécanismes impliqués dans l'augmentation d'expression de l'ATX avec la différenciation adipocytaire restent à clarifier.

L'expression de l'ATX a été mesurée dans le tissu adipeux de divers modèles murins d'obésité. Parmi les différents modèles étudiés (souris engraissées par un régime hyperlipidique, souris traitées à l'aurothioglucose, souris db/db) seules les souris db/db présentent une augmentation importante de l'expression de l'ATX adipocytaire (Ferry et al 2003 ; Boucher et al 2005). Ces souris se distinguent des autres modèles d'obésité par une hyper-insulinémie supérieure associée à une forte diminution de la sensibilité à l'insuline de ses adipocytes. Chez l'Homme, aucune corrélation significative n'a été retrouvée entre l'indice de masse corporelle (IMC) et l'expression de l'ATX du tissu adipeux. Par contre, une augmentation significative d'expression de l'ATX a été observée dans le tissu adipeux de patientes associant obésité massive (IMC supérieur à 40) et intolérance au glucose (Boucher et al 2005). L'ensemble de ces observations montrent que l'expression de l'ATX du tissu adipeux n'est pas directement liée à l'état d'engraissement des individus, mais pourrait être dépendante de l'état de résistance à l'insuline des adipocytes et/ou à l'inflammation du tissu adipeux.

**6/ Conclusions - Perspectives.**

Ces études montrent que l'ATX, par l'intermédiaire du LPA et du récepteur LPA1, pourrait participer à la régulation paracrine de l'adipogénèse et/ou à l'étiologie du diabète associé à l'obésité. Cependant, des études complémentaires doivent être menées pour vérifier ces hypothèses. Il sera nécessaire de parfaire nos connaissances des mécanismes cellulaires et moléculaires impliqués dans les changements d'expression de l'ATX et du récepteur LPA1 au cours de la différenciation adipocytaire ainsi que dans le diabète associé à l'obésité. Il sera également nécessaire de développer des modèles de souris transgéniques où l'expression de l'ATX et du récepteur LPA1 du tissu adipeux sont modifiées (invalidation, sur-expression) et d'en déterminer les conséquences phénotypiques. Si nos hypothèses sont vérifiées, l'ATX et le récepteur LPA1 pourraient constituer des cibles pharmacologiques intéressantes dans le traitement de l'obésité et/ou de certaines pathologies associées comme le diabète.